\documentclass{aa}

\usepackage{graphicx}
\usepackage{txfonts}

\usepackage{amsmath}    
\usepackage{amssymb}    
\usepackage{mathtools}

\usepackage{threeparttable}
\usepackage{longtable}
\usepackage{tabto}
\usepackage{adjustbox}
\usepackage{multicol} 
\usepackage{multirow}

\usepackage{color}
\usepackage{natbib,twoopt}
\usepackage[hyphenbreaks]{breakurl}
\usepackage[breaklinks]{hyperref}      
\bibpunct{(}{)}{;}{a}{}{,}             
\definecolor{cobalt}{rgb}{0.06, 0.2, 0.65}
\hypersetup{
  colorlinks,
  citecolor=blue,
  linkcolor=blue,
  urlcolor=blue,
}
\makeatletter
  \newcommandtwoopt{\citeads}[3][][]{\href{http://adsabs.harvard.edu/abs/#3}%
    {\def\hyper@linkstart##1##2{}%
     \let\hyper@linkend\@empty\citealp[#1][#2]{#3}}}
  \newcommandtwoopt{\citepads}[3][][]{\href{http://adsabs.harvard.edu/abs/#3}%
    {\def\hyper@linkstart##1##2{}%
     \let\hyper@linkend\@empty\citep[#1][#2]{#3}}}
  \newcommandtwoopt{\citetads}[3][][]{\href{http://adsabs.harvard.edu/abs/#3}%
    {\def\hyper@linkstart##1##2{}%
     \let\hyper@linkend\@empty\citet[#1][#2]{#3}}}
  \newcommandtwoopt{\citeyearads}[3][][]%
    {\href{http://adsabs.harvard.edu/abs/#3}
    {\def\hyper@linkstart##1##2{}%
     \let\hyper@linkend\@empty\citeyear[#1][#2]{#3}}}
\makeatother

\newcommand{\Msun}{M$_{\odot}$}
\newcommand{\Msunyr}{M$_{\odot}$~yr$^{-1}$}
\newcommand{\Rsun}{R$_{\odot}$}

\definecolor{smalt(darkpowderblue)}{rgb}{0.0, 0.2, 0.6}
\definecolor{forestgreen(traditional)}{rgb}{0.0, 0.5, 0.0}

\defcitealias{RVJ}{RVJ}

\defcitealias{MD17}{MD17}

\newcommand{\cv}{cataclysmic variable}
\newcommand{\cvs}{cataclysmic variables}
\newcommand{\Cv}{Cataclysmic variable}

\newcommand{\elm}{extremely low-mass}

\newcommand{\agb}{asymptotic giant branch}

\newcommand{\paradoxical}{SDSS\,J1257$+$5428}

\newcommand{\bse}{BSE}

\newcommand{\mesa}{MESA}

\usepackage[dvipsnames]{xcolor}

\begin{document}

   \title{Resolution of a paradox: SDSS~J1257+5428 can be explained as a descendant of a cataclysmic variable with an evolved donor}

   \titlerunning{Explaining SDSS~J1257+5428 as a descendant of a CV with an evolved donor}

    \author{Diogo Belloni
           \inst{1,2}
    \and
          Matthias R. Schreiber 
          \inst{1}
    \and
          Kareem El-Badry
          \inst{3}
          }

    \authorrunning{D. Belloni et al.}

    \institute{Departamento de F\'isica, Universidad 
               T\'ecnica Federico Santa Mar\'ia, Av. España 1680, Valpara\'iso, Chile\\
              \email{diogobellonizorzi@gmail.com}
    \and
              São Paulo State University (UNESP), School of Engineering and Sciences, Guaratinguetá, Brazil
    \and
              Department of Astronomy, California Institute of Technology, 1200 E. California Blvd., Pasadena, CA 91125, USA
             }

   \date{Received...; accepted ...}

 
  \abstract
   {
   The binary system SDSS\,J1257+5428 consists of an extremely low-mass white dwarf (WD) with a mass ranging from ${\sim0.1}$ to ${\sim0.24}$~\Msun, along with a more massive WD companion of approximately $1$~\Msun\ that is significantly hotter. Recently, a tertiary WD orbiting this binary was discovered, setting a lower limit for the total age (${\sim4}$~Gyr) of the triple and providing further constraints on SDSS\,J1257+5428 that could be used to constrain its formation pathways. Up to now, the existence of this system has been described as paradoxical since tested models for its formation cannot account for its properties.
   }
   {
   Here we investigate under which conditions SDSS\,J1257+5428 could be understood as a descendant of a cataclysmic variable with an evolved donor star, which is a scenario that has never been explored in detail.
   }
   {
   We used the rapid BSE code for pre-common-envelope (CE) evolution and the detailed MESA code for post-CE evolution to carry out binary evolution simulations and searched for potential formation pathways for SDSS\,J1257+5428 that lead to its observed characteristics. For the post-CE evolution, we adopted a boosted version of the CARB model, which has been successfully used to explain the properties of close binary stars hosting evolved solar-type stars.
   }
   {
   We find that SDSS\,J1257+5428 can be explained as a post-cataclysmic-variable system if (i) the progenitor of the extremely low-mass WD was initially a solar-type star that evolved into a subgiant before the onset of mass transfer and underwent hydrogen shell flashes after the mass transfer stopped, (ii) the massive WD was highly or entirely rejuvenated during the cataclysmic variable evolution, and (iii) magnetic braking was strong enough to make the evolution convergent. In this case, the torques due to magnetic braking need to be stronger than those provided by the CARB model by a factor of ${\sim100}$.
   }
   {
   We conclude that SDSS\,J1257+5428 can be reasonably well explained as having originated from a cataclysmic variable that hosted an evolved donor star and should no longer be regarded as paradoxical. If our formation channel is correct, our findings provide further support that stronger magnetic braking acts on progenitors of (i) close detached WD binaries, (ii) close detached millisecond pulsar with extremely low-mass WDs, (iii) AM\,CVn binaries, and (iv) ultra-compact X-ray binaries, in comparison to the magnetic braking strength required to explain binaries hosting main-sequence stars and single main-sequence stars.
   }


   \maketitle
%


\section{Introduction} 
\label{Introduction}

Binary stars that undergo episodes of mass transfer leading to pairs of white dwarfs (WDs) with small separations are likely the most abundant population among Type Ia supernova progenitors \citep[see][for a recent review]{Liu_2023}.
In addition, they are expected to be the main source of detectable low-frequency gravitational waves \citep[e.g.,][]{Korol_2020,AmaroSeoane_2023,LiChen_2024}.
Despite their importance, we still struggle to understand the mechanisms leading to their formation \citep[see][for a recent review]{BelloniSchreiberREVIEW}.

Among double WD binaries, SDSS~J125733.63$+$542850.5 (hereafter \paradoxical) deserves special attention since its formation pathway has been a mystery for years.
It was discovered by \citet{Badenes_2009} as a circular, $4.56$-hour period, single-lined, spectroscopic binary from SWARMS (Sloan White dwArf Radial velocity data Mining Survey).
From the spectroscopic and photometric data, these authors estimated a mass of ${0.92^{+0.28}_{-0.32}}$~\Msun~for one of the components and concluded that the unseen companion must be more massive than ${1.62^{+0.20}_{-0.25}}$~\Msun, suggesting that it is most likely a neutron star.
As shown later by \citet{KulkarnivanKerkwijk_2010} and \citet{Marsh_2011}, follow-up B- and R-band spectroscopy confirmed that the system is instead composed of two WDs, a cool \elm~WD and a hotter massive WD.

Subsequently, \citet{Bours_2015} gathered ultraviolet data with the \textit{Hubble} Space Telescope and refined the parameters of both WDs.
They found that the cold component has a mass of ${\sim0.1-0.24}$~\Msun~and an effective temperature of ${\sim6\,400}$~K, while its hot companion has a mass of ${1.06\pm0.05}$~\Msun~and an effective temperature of ${\sim13\,030}$~K.
If the properties of the first WD formed are unaffected by the formation process of the second WD (something that cannot be confirmed by observations), then these masses and effective temperatures imply that the \elm~WD is significantly older than the massive WD.
Because of these apparently contradicting properties, the existence of \paradoxical~has been treated as a paradox for binary star evolution.

Most recently, \citet{ArosBunster_2025} inspected the \textit{Gaia} Data Release 3 archive and discovered that the object 1570271757456852096 is a common proper motion WD companion, making \paradoxical~the third known triple WD system.
By using Keck Low Resolution Imaging Spectrometer spectroscopy, \textit{Gaia} and Sloan Digital Sky Survey (SDSS) photometry, and WD atmosphere models, these authors estimated a cooling age of ${\gtrsim4}$~Gyr for the tertiary, irrespective of the cooling model and its atmospheric composition.
This then imposes strong constraints on the total age of the triple, regardless of whether the tertiary originated from single star evolution or from binary star evolution through a merger.

The age of the tertiary would not be a real constraint if it were not bound to the binary initially, that is, if the triple were formed dynamically through a three- or four-body interaction.
As shown by \citet{HeggieBOOK}, the probability of triple formation through tidal capture in binary-single interactions is negligible.
On the other hand, the probability of hierarchical triples forming through binary-binary interactions might be significant, according to numerical scattering experiments \citep[e.g.,][]{Mikkola_1984,Leigh_2016b,Ryu_2017,BarreraRetamal_2024}.
Despite that, dynamical formation in the case of \paradoxical~seems very unlikely.
The system is at a distance of $120$~pc and belongs to the Galactic disk, where four-body interactions are not expected to happen as the disk is not dense enough for stellar encounters to play an important role \citep[e.g.,][]{BinneyBOOK}.

The present-day small orbital separations of close double WD binaries imply that these systems are currently much smaller in comparison to when they host a red giant (the progenitor of at least one of the WDs). 
Therefore, the orbit of these binaries must have been significantly reduced during their formation, most likely through common-envelope (CE) evolution triggered by dynamically unstable mass transfer.
Although one of the required episodes of mass transfer was likely CE evolution, the other required episode could in principle be either CE evolution or dynamically stable mass transfer.
Regarding the possible formation channels for \paradoxical, there are in principle three possibilities, all involving two episodes of mass transfer, each forming one WD: (i) two episodes of CE evolution to form the hot and cold components, (ii) dynamically stable mass transfer (to form the cold component) followed by CE evolution (to form the hot component), and (iii) CE evolution (to form the hot component) followed by dynamically stable mass transfer (to form the cold component).

The first scenario can be ruled out because the massive and hot WD would form first as it originates from an initially more massive star, which would mean that it should be cooler than observed.
The mass of the progenitor of the massive WD (${\sim1}$~\Msun) was initially ${\sim5}$~\Msun~\citep[e.g.,][]{Cummings_2018}, and it quickly evolved into a WD (${\sim100}$~Myr).
The \elm~WD companion must have originated from a star with an initial mass of ${\lesssim2}$~\Msun~as otherwise the helium core after the hydrogen-burning phase (lasting ${\gtrsim1}$~Gyr) would be more massive than estimated from observations.
This implies that the age of the massive WD must have been ${\gtrsim1}$~Gyr when the \elm~proto-WD was formed.
At the present day, the \elm~WD has an age of ${\gtrsim1.0}$~Gyr, which would imply an age of ${\gtrsim2}$~Gyr and a temperature of ${\lesssim10\,000}$~K \citep{Bedard_2020} for the massive WD. This is much lower than observed (${\sim13\,000}$~K).
Another piece of evidence against this scenario comes from the total age of the triple.
As it is ${\gtrsim4}$~Gyr, the present-day temperature of the massive WD should be ${\lesssim8\,500}$~K \citep{Bedard_2020}, which is much lower than observed.

The second scenario was recently investigated by \citet{ArosBunster_2025}.
In this case, the cold, lower-mass WD is assumed to form first, as in principle this is consistent with the constraints from observations, which suggests that the cold, lower-mass WD is older than the hot component.
However, the first episode of mass transfer would lead to a binary with a much shorter orbital period than what is required to accommodate the progenitor of a WD with a mass of ${\sim1}$~\Msun.
Even if this difficulty could somehow be overcome, 
\citet{ArosBunster_2025} show that the age difference between the cold and hot components would only correspond to ${\sim100}$~Myr, which is much shorter than estimated from observations.

Therefore, the only alternative left is the third scenario, that is, the hot WD  formed first through CE evolution, and the cold component formed afterward, due to dynamically stable mass transfer (i.e., cataclysmic variable evolution).
Although at first glance it may seem that this scenario should not even be considered because the hot WD would be colder if formed first, this is not necessarily true if the properties of the hot WD change during the formation of the cold WD.
In this work we ran binary evolution models to investigate whether and under which conditions this third scenario, the only remaining option, could lead to the resolution of the paradox.
We find that this scenario is indeed possible provided the donor star is a subgiant that later underwent hydrogen shell flashes as a proto-WD, the hot component is highly or entirely rejuvenated during cataclysmic variable evolution, and the magnetic braking is sufficiently strong to make the cataclysmic variable evolution convergent.

\begin{figure*}
\begin{center}
\includegraphics[width=0.99\linewidth]{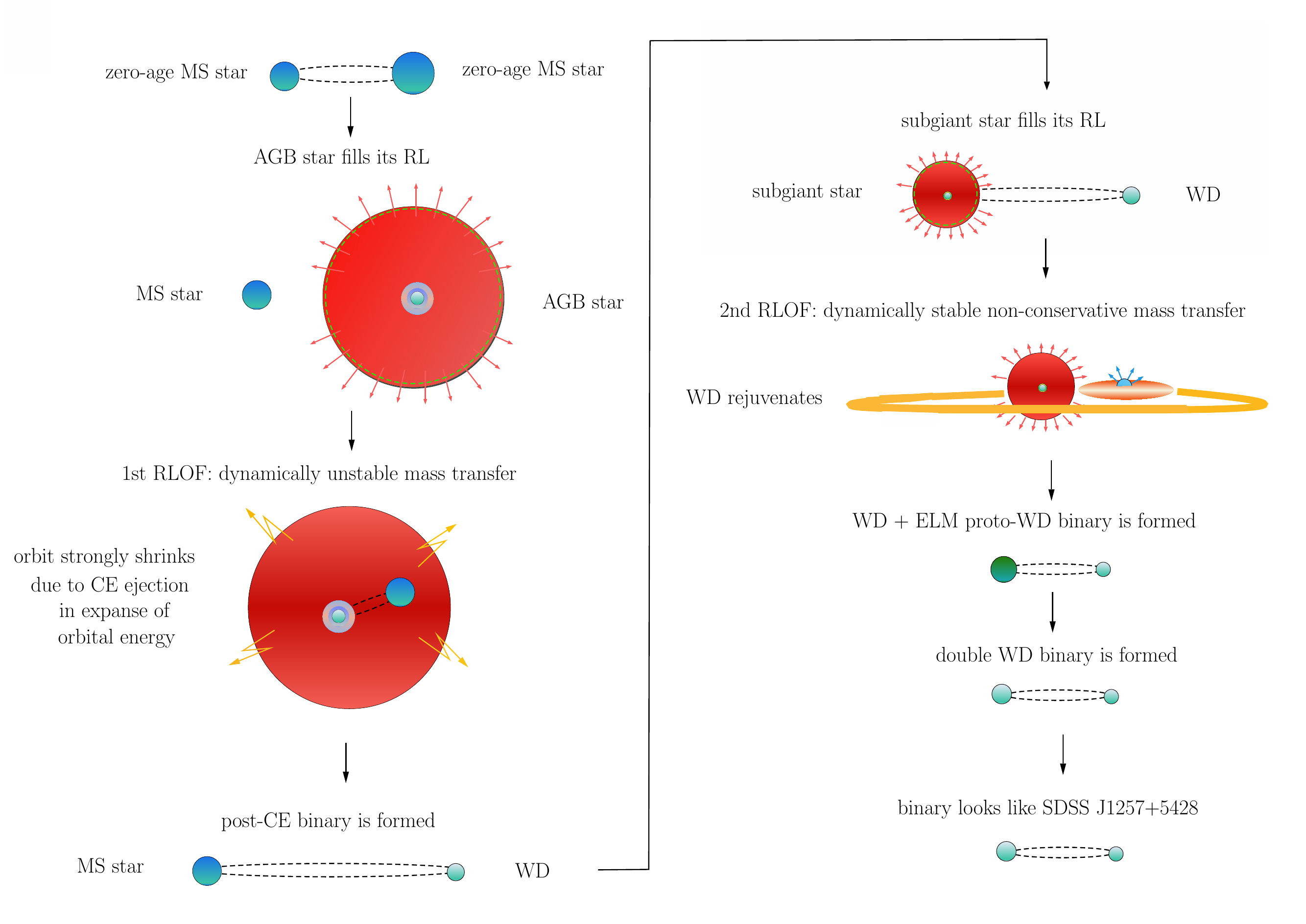}
\end{center}
\caption{Proposed formation pathway for \paradoxical. After the more massive star in the zero-age main-sequence binary becomes an AGB star and fills its Roche lobe, the binary evolves through a CE phase. The resulting post-CE binary consists of the newly formed WD and a main-sequence companion star. The orbital period of the post-CE binary is such that the WD has time to cool and the main-sequence star has time to evolve and become a subgiant before filling its Roche lobe. During \cv~evolution, the WD is heated due to accretion and rejuvenated. Following \cv~evolution, the donor becomes an \elm~proto-WD and after undergoing hydrogen flashes finally becomes an \elm~WD. The binary eventually looks like \paradoxical.}
\label{Fig:Schematic}
\end{figure*}

\section{A testable formation channel for \paradoxical}

In the scenario we investigate here, the WD that becomes the massive, hot WD was formed first via CE evolution, followed by the formation of the WD that becomes the \elm, cold WD via dynamically stable mass transfer.
In what follows, we provide an overview of the general concepts on which this formation channel is based.

After the more massive star of the initial main-sequence binary system evolves off the main sequence, it becomes an \agb~(AGB) star and eventually fills its Roche lobe.
Provided this Roche-lobe-filling red giant cannot maintain hydrostatic equilibrium, the system enters a phase of dynamically unstable mass transfer. 
In such a case, the mass-loss timescale becomes extremely short, preventing the donor star from remaining within its Roche lobe, which results in the formation of a CE surrounding the dense giant core and the main-sequence star.
This triggers a rapid orbital decay and ultimately leads to an episode of CE evolution and the formation of the WD that later becomes the hot WD in \paradoxical.
If the orbital period of the resulting post-CE binary is sufficiently long (more than a few days), the companion of the WD will have enough time to evolve off the main sequence and become a subgiant, before filling its Roche lobe.

After CE evolution, the subsequent evolution of the binary, however, is not only driven by nuclear evolution but also by orbital angular momentum loss due to magnetic wind braking.
The strength of magnetic braking and its dependence on stellar mass, stellar structure, and evolutionary stage are highly uncertain.
However, there is growing evidence that for binary stars containing main-sequence star companions, magnetic braking is efficient for stars with a radiative core and a convective envelope but inefficient for fully convective stars and stars with a convective core \citep{schreiberetal10-1,Knigge_2011_OK,Zorotovic_2017,BelloniSchreiberREVIEW,Belloni_2024a,Schreiber_2024}.

While magnetic braking is even less constrained for slightly evolved stars, based on the above, it is likely that if the companion star has a radiative core, the evolution of the binary consisting of the WD and its evolved companion is driven by a combination of orbital angular momentum loss due to magnetic braking and mass loss through winds.
On the other hand, for stars having convective cores, magnetic braking is likely inefficient and the evolution is driven by the mass loss through winds only.
If the companion of the WD was a main-sequence star with a mass of ${\gtrsim1.2}$~\Msun, then its core became radiative only after it became a subgiant and only at this point magnetic braking became the driver of the binary evolution.
Alternatively, if its mass was smaller than that, then magnetic braking is always acting since the core of the star is radiative.

Regardless of when magnetic braking kicked in, the donor star has to be a subgiant at the onset of mass transfer and the evolution has to be driven by strong orbital angular momentum loss due to magnetic braking.
Only if at the onset of mass transfer magnetic braking dominates over the nuclear evolution of the donor star, the system evolves toward shorter orbital periods, which is required to explain the properties of \paradoxical.
As soon as the donor star fills its Roche lobe, dynamically stable mass transfer (i.e., \cv~evolution) starts.
Assuming that magnetic braking dominates, the orbital period and the donor mass decrease as a consequence of the orbital angular momentum loss.
When the mass transfer rate is sufficiently high, hydrogen can be stably burnt on the WD surface, which leads to an increase in its mass and an overall rejuvenation.
As soon as the mass of the donor envelope becomes negligible, the mass transfer rate (due to magnetic braking and mass loss) drops and the binary detaches.
At this point, mass transfer ends and the remaining binary consists of a massive WD paired with an \elm~proto-WD \citep[see][for more details]{BelloniSchreiber_2023}.

We would like to draw the readers attention to the important differences between the detachment phase in standard \cv~evolution and the detachment phase leading to the formation of double WD binaries.
In the former, the donor star is an unevolved main-sequence star that transitions to a fully convective main-sequence star when its mass drops to ${\sim0.2}$~\Msun~\citep{McAllister_2019}.
Although magnetic braking is not fully disrupted \citep[e.g.,][]{ElBadry_2022,Belloni_2024a}, angular momentum loss through magnetic braking becomes less efficient, allowing the donor to detach from its Roche lobe.
On the other hand, when the donor star is a subgiant, its core is radiative and remains as such during the entire \cv~evolution.
When the mass of the envelope of the donor star becomes sufficiently low, hydrogen shell burning can no longer prevent the contraction of the donor star.  This contraction causes the donor star to detach from its Roche lobe and to evolve into an \elm~proto-WD.

In the emerging double degenerate binary, the age of the massive WD is reset during \cv~evolution. 
The companion, an \elm~proto-WD, is sufficiently massive to evolve through hydrogen shell flashes.
Subsequently, the \elm~WD enters its cooling sequence and from this point on the binary evolution is driven solely by the emission of gravitational waves, which causes the orbital period to decrease smoothly.
The resulting system stellar and binary parameters may then be comparable to those of \paradoxical.
Throughout the remainder of this paper, we test under which conditions this is the case.
A schematic of our proposed scenario is provided in Fig.~\ref{Fig:Schematic}.

\section{Binary evolution models}
\label{Methods}

We tested the hypothesis that \paradoxical~is a descendant of a \cv~with an evolved donor adopting a two-code scheme.
For pre-CE and CE evolution we used the BSE code \citep{Hurley_2000,Hurley_2002}, which is a rapid population synthesis code that allows us to quickly explore a broad region of the parameter space 
\citep[e.g.,][]{Belloni_2024b}.
For post-CE evolution, we used the MESA code because it solves the fully coupled structure and composition equations simultaneously and is capable of calculating fully self-consistent evolutionary tracks through stable mass transfer. The results of these evolutionary tracks can be compared with the characteristics of \paradoxical.
In what follows, we describe in detail our assumptions for pre-CE, CE and post-CE evolution.

\subsection{Pre-common-envelope evolution modeling}

Unless clearly stated otherwise, we used the standard \bse~values for all stellar and binary evolution parameters \citep[e.g.,][]{Hurley_2002}.
Our approach follows closely that of \citet{Belloni_2024b}.

We assumed solar metallicity (i.e., ${Z=0.02}$) and set the \citet{Reimers_1975} wind efficiency to $0.5$ for the first giant branch evolution.
During the evolution on the AGB, we adopted the prescription proposed by \citet{VW93}.
Regarding wind accretion, we adopted the Bondi-Hoyle-Lyttleton prescription \citep{Hoyle_1939,Bondi_1944}.

For CE evolution, we adopted the energy formalism, that is, the binding energy of the envelope of the red giant donor ($E_\mathrm{bind}$) at the onset of the CE evolution is assumed to be equal to the change in orbital energy during the spiral-in phase ($\Delta E_\mathrm{orb}$) scaled with a parameter $\alpha_{\rm CE}$, corresponding to the fraction of the change in orbital energy that is used to unbind the envelope:

\begin{equation}
E_{\rm bind} \ = \ 
\alpha_{\rm CE} \ \Delta E_{\rm orb} \ = \ - \ 
\alpha_{\rm CE} \
 \left( \, \frac{G\,M_{\rm d,c}\,M_{\rm a}}{2\,a_f}  \ - \ 
           \frac{G\,M_{\rm d,c}\,M_{\rm a}}{2\,a_i} \, \right) \ ,
\label{EQALPHACE}
\end{equation}

\noindent
where $M_{\rm a}$ is the accretor mass, $M_{\rm d,c}$ is the core mass of the donor, $a_i$ is the semimajor axis at the onset of the CE evolution, and $a_f$ is the semimajor axis after CE ejection.

The binding energy is approximated by
\begin{equation}
E_{\rm bind} \ = \ - \ \frac{G\,M_{\rm d}\,(M_{\rm d}\,-\,M_{\rm d,c})}{\lambda \, R_{\rm d}} \ ,
\label{EQLAMBDA}
\end{equation}

\noindent
where $M_{\rm d}$ is the donor mass, $R_{\rm d}$ is the donor radius, and $\lambda$ is the envelope-structure parameter
\citep{Dewi_2000,Xu_2010,Loveridge_2011,Klencki_2021,Marchant_2021}.

We set the CE efficiency to ${\alpha=0.3}$, which is consistent with the increasing evidence that short-period post-CE binary progenitors experience strong orbital shrinkage during CE evolution \citep[e.g.,][]{Zorotovic_2010,Toonen_2013,Camacho_2014,Cojocaru_2017,Belloni_2019,Hernandez_2022a,Zorotovic_2022,Scherbak_2023,Chen_2024}.
The envelope-structure parameter $\lambda$ was calculated according to a fitting scheme similar to that provided by \citet[][their Appendix A]{Claeys_2014}, which is based on the detailed numerical stellar evolution calculations by \citet{Dewi_2000} and takes into account the structure and the evolutionary stage of the red giant donor and the envelope thermal energy as constrained by the virial theorem (i.e., increasing $\lambda$ by a factor of 2).

\subsection{Post-common-envelope evolution modeling}

We used the version r15140 of the MESA code \citep[][]{Paxton2011,Paxton2013,Paxton2015,Paxton2018,Paxton2019,Jermyn2023} to calculate binary evolution after CE evolution.
For reference, our approach follows closely that by \citet{BelloniSchreiber_2023}\,\footnote{\href{https://zenodo.org/records/8279474}{https://zenodo.org/records/8279474}}, \citet{Belloni_2024c}\,\footnote{\href{https://zenodo.org/records/10841636}{https://zenodo.org/records/10841636}}, and \citet{Belloni_2024d}\,\footnote{\href{https://zenodo.org/records/10937460}{https://zenodo.org/records/10937460}}.

\subsubsection{Stellar evolution} 
\label{MESAassumptionsStar}

The MESA equation of state is a blend of the OPAL \citep{Rogers2002}, SCVH \citep{Saumon1995}, FreeEOS \citep{Irwin2004}, HELM \citep{Timmes2000}, PC \citep{Potekhin2010} and Skye \citep{Jermyn2021} equations of state.
Nuclear reaction rates are a combination of rates from NACRE \citep{Angulo1999}, JINA REACLIB \citep{Cyburt2010}, plus additional tabulated weak reaction rates \citep{Fuller1985,Oda1994,Langanke2000}.
Screening is included via the prescription of \citet{Chugunov2007} and thermal neutrino loss rates are from \citet{Itoh1996}.
Electron conduction opacities are from \citet{Cassisi2007} and radiative opacities are primarily from OPAL \citep{Iglesias1993,Iglesias1996}, with high-temperature Compton-scattering dominated regime calculated using the equations of \citet{Buchler1976}.

We adopted a metallicity of ${Z=0.02}$ and the grey Eddington T(tau) relation to calculate the outer boundary conditions of the atmosphere, using a uniform opacity that is iterated to be consistent with the final surface temperature and pressure at the base of the atmosphere.
Radiative opacities are taken from \citet{Ferguson2005} for $2.7\leq \log T \leq4.5$ and OPAL \citep{Iglesias1993,Iglesias1996} for $3.75\leq \log T \leq8.7$.
For the evolutionary phases with convective core, that is, core hydrogen and helium burning, we took into account exponential diffusive overshooting, assuming a smooth transition in the range ${1.2-2.0}$~\Msun~\citep[e.g.,][]{Anders_2023}.
We assumed that the extent of the overshoot region corresponds to a fraction of ${H_{\rm p}}$ \citep[e.g.,][]{Schaller_1992,Freytag_1996,Herwig_2000}, with $H_{\rm p}$ being the pressure scale height at the convective boundary.
We further used the nuclear network \texttt{cno$\_$extras.net}, which accounts for the nuclear reactions of the carbon-nitrogen-oxygen cycle for hydrogen burning.

We allowed the star to lose mass through winds, adopting the \citet{Reimers_1975} prescription and setting the wind efficiency to $0.5$.
We treated convective regions using the scheme by \citet{Henyey_1965} for the mixing-length theory, assuming that the mixing length is a factor of ${H_{\rm p}}$ \citep[e.g.,][]{Joyce_2023}.
In addition, the boundaries of convective regions are determined using the Schwarzschild criterion.
We also included element diffusion due to gravitational settling and chemical and thermal diffusion \citep{Paxton2015} for $^1$H, $^{3}$He, $^{4}$He, $^{12}$C, $^{13}$C, $^{14}$N, $^{16}$O, $^{20}$N, $^{24}$Mg, and $^{40}$Ca.

Regarding the WD accretor, we approximated it with a point mass.
In addition, we assumed that it was entirely rejuvenated during \cv~evolution, that is, its age was reset to zero at the onset of the detachment (i.e., when the donor becomes a proto-WD), which is consistent with simple prescriptions for the cooling and heating of accreting WDs \citep{Schreiber_2023}.
The effective temperature and gravity of the cooling massive WD in the detached double degenerate binary were determined via interpolating the WD evolutionary models with thick atmospheres (surface hydrogen abundance of $10^{-4}$) computed by \citet{Bedard_2020}.

\subsubsection{Binary evolution}
\label{MESAassumptionsBinary}

The Roche-lobe radius of each star was computed using the fit of \citet{Eggleton1983}.
The mass transfer rates due to Roche-lobe overflow are determined following the prescription of \citet{Ritter1988}, in which the atmosphere of the star is filling the Roche lobe.
In this so-called atmospheric Roche-lobe overflow model, mass transfer occurs even when the star radius is smaller than the Roche-lobe radius.

Mass transfer due to Roche-lobe overflow is likely non-conservative.
We allowed mass and angular momentum losses following the formalism of \citet{SPV97}.
The mass lost from the vicinity of the donor star as a fast wind (i.e., Jeans mode) corresponds to a fraction $\alpha$ of the transferred mass and is modeled as a spherically symmetric outflow from the donor star in the form of a fast wind.
The mass lost from the vicinity of the WD accretor as a fast wind (i.e., isotropic re-emission) corresponds to a fraction $\beta$ of the accreted mass.
Finally, the mass lost from a circumbinary disk corresponds to a fraction $\delta$ of the transferred mass and is modeled as a time-dependent coplanar toroid (see below).
Thus, the accretion efficiency during mass transfer is then given by
\begin{equation}
\epsilon = 1 - \alpha - \beta - \delta.
\end{equation}

As we just mentioned, we assumed that the WD accretor can only accrete a part of the mass transferred from its companion.
Depending on the accretion rate onto the WD, hydrogen shell burning can be stable resulting in an increase in its mass.
This has been modeled by implementing the critical accretion rate calculated by \citet[][]{Wolf_2013}, above which stable hydrogen shell burning occurs. 
For accretion rates lower than this critical value, the WD undergoes nova eruptions, such that almost all of the accreted mass is expelled from the binary.
Since it is unlikely that exactly the entire accreted mass is ejected \citep[e.g.,][]{Jose_2020}, we assumed that ${\beta=85\%}$.
We further assumed that there is a maximum possible accretion rate \citep[][]{Wolf_2013} such that WDs accreting at higher rates will burn stably at this maximum rate and the remaining non-accreted matter will be piled up forming a red-giant-like envelope, which is assumed to be lost from the binary in the form of stellar-like fast winds.

For the orbital angular momentum loss, we took into account not only those related to the mass loss from the system but also due to gravitational radiation and magnetic braking as
\begin{equation}
  \frac{\dot{J}_{\rm orb}}{J_{\rm orb}} = 
         \frac{\dot{J}_{\rm ML}}{J_{\rm orb}}                         
       + \frac{\dot{J}_{\rm CB}}{J_{\rm orb}}                         
       + \frac{\dot{J}_{\rm GR}}{J_{\rm orb}}
       + \frac{\dot{J}_{\rm MB}}{J_{\rm orb}} \,,
  \label{Jdot}
\end{equation}
\noindent
where
\begin{eqnarray}
J_{\rm orb} \ = \ M_{\rm WD} \, M_2 \, \sqrt{\frac{G \, a}{ M_{\rm WD} \ + \ M_2 }}\,,
\label{J_orb}
\end{eqnarray}
\noindent
where $G$ is the gravitational constant, $M_{\rm WD}$ is the WD mass, $M_2$ is the donor mass, and $a$ is the orbital separation.

The fraction of accreted mass leaving the WD is assumed to carry away the specific angular momentum of the WD and the mass lost from the donor star carries the specific angular momentum of the donor star.
In particular, the orbital angular momentum loss due to mass loss is given by
\begin{equation}
\frac{\dot{J}_{\rm ML}}{J_{\rm orb}}
\ = \ 
\left(\,
  \beta\,\frac{q^2}{1\,+\,q}
  \ + \
  \alpha\,\frac{1}{1\,+\,q}\,
\right)
\,\frac{\dot{M}_2}{M_2}  \,,
\label{J_ML}
\end{equation}
\noindent
where $q=M_2/M_{\rm WD}$ is the ratio between donor star mass and the mass of the WD accretor.

For the orbital angular momentum loss due to a circumbinary disk, following \citet{SpruitTaam_2001}, we assumed that a fraction of the mass transfer rate feeds a circumbinary disk (i.e., is lost through the outer Lagrange point) and the circumbinary disk is modeled as a time-dependent toroid with inner radius $r_i$ corresponding to $1.7$ times the binary separation, that is,
\begin{equation}
\frac{\dot{J}_{\rm CB}}{J_{\rm orb}}
\ = \ 
\gamma\,\delta\,\left(1\,+\,q\right)
\,\frac{\dot{M}_2}{M_2}  \,,
\label{J_CB}
\end{equation}
\noindent
where
\begin{equation}
\gamma \ = \ 1.3 \,
\left( \, r_i/a \, \right)^{1/2} \, 
\left( \, t_* / t_{\rm vi} \, \right)^{1/2} \,,
\label{J_CB_gamma}
\end{equation}
\noindent
${\delta=0.1\%}$, $t_{\rm vi}=100$~yr is the disk viscous timescale, and $t_*$ is the time in yr since the onset of mass transfer.
The binary then has initially an empty circumbinary disk, which is fed with a rate corresponding to a fraction $\delta$ of the mass transfer rate.
This progressive buildup of the disk mass implies a growing torque during the \cv~evolution.

The orbital decay due to the emission of gravitational waves is given by
\begin{equation}
\dot{J}_{\rm GR} = -\frac{32 }{5c^5}\left(\frac{2\pi G}{P_{\rm orb}}\right)^{7/3}\frac{(M_1M_2)^2}{(M_1+M_2)^{2/3}}.
\label{J_GR}
\end{equation}
For magnetic braking, we adopted a boosted version of the CARB  prescription \citep{CARB} given by
\begin{equation}
\begin{multlined}
\dot{J}_{\rm MB} \ = \ - \, \eta \times \
\left( 2\times10^{-6} \right)
\left(
   \frac{-\dot{M}_{\rm wind}}{\rm g~s^{-1}}
\right)^{-1/3}
\left(
   \frac{R_2}{\rm cm}
\right)^{14/3}
\left(
   \frac{\Omega_2}{\Omega_\odot}
\right)^{11/3} \, \times
\\
\left(
    \frac{\tau_{\rm conv} }{\tau_{\odot, \rm conv}}
\right)^{8/3}
\left[
  \left(
      \frac{v_{\rm esc}}{\rm cm~s^{-1}}
  \right)^2\,+\,\frac{2}{K_2^2}\,
  \left(
      \frac{\Omega_2}{\rm s^{-1}}
  \right)^2
  \left(
      \frac{R_2}{\rm cm}
  \right)^2
\right]^{-2/3} \, ,
\end{multlined}
\label{J_MB}
\end{equation}

\noindent
where $\dot{M}_{\rm wind}$, $R_2$, $\Omega_2$, and $\tau_{\rm conv}$ are the wind mass-loss rate, radius, spin and convective turnover timescale of the companion of the WD, respectively.
We introduced the boosting parameter ${\eta\geq1}$ so that magnetic braking is allowed to be stronger than predicted by the standard CARB model.
The convective turnover timescale was calculated by integrating the inverse of the velocity of convective cells, as given by the mixing-length theory, over the radial extent of the convective envelope following \citet{CARB}.
The spin of the Sun and its convective turnover timescale are ${3\times10^{-6}}$~${\rm s^{-1}}$ and ${2.8\times10^6}$~s, respectively, and ${K_2=0.07}$.
Finally, $v_{\rm esc}$ is the escape velocity.

We enabled magnetic braking only during the evolutionary phases in which both the convective envelope and the radiative core are non-negligible.
During \cv~evolution, when the mass fraction of the convective envelope becomes less than 2\%, the strength of magnetic braking is reduced by a factor of $e^{1-0.02/q_{\rm conv}}$ \citep{Podsiadlowski_2002}, where $q_{\rm conv}$ is the mass fraction of the convective envelope.
By doing that, we assumed that donors with very little mass in their convective envelopes do not generate strong magnetic fields and, as a result, experience minimal magnetic braking.

\section{Searching for a reasonable binary model for \paradoxical} 
\label{Search}

Based on the assumptions described in Sect.~\ref{Methods}, our search for an evolutionary pathway leading to a system as similar as possible to \paradoxical~was performed in two steps.
First, we searched for the best-fitting post-CE model with the MESA code by varying several stellar/binary evolution parameters as well as the initial post-CE binary parameters.
Second, once the best-fitting post-CE model was found, we searched with the BSE code for the zero-age binary that evolves to an initial post-CE binary with the same properties we found in the previous steps.
We describe below in more detail our fitting scheme.

In our search for the best-fitting post-CE model, we ran different grids of models.
Upon finding the model within the grid that was nearest to the observed one, we refined the grid around this model to ultimately achieve an optimal fitting.
The observed parameters we used to constrain the models are the (i) the mass, effective temperature and $\log{\rm g}$ of the \elm~WD (i.e., ${\lesssim0.24}$~\Msun, ${\sim6\,300-6\,600}$~K, ${\sim6-7}$, respectively); (ii) the mass, effective temperature and $\log{\rm g}$ of the massive WD (i.e., ${1.06\pm0.05}$~\Msun, ${\sim12\,600-13\,200}$~K, ${\sim8.5-8.8}$, respectively); (iii) the orbital period (i.e., $4.56$~hr); (iv) the age of the massive WD (i.e., ${\gtrsim4}$~Gyr), and (v) the ratio between the radii of the extremely low-mass WD and the massive WD (i.e., ${\sim3.8-4.4}$).

In our fitting scheme, we varied the initial post-CE WD mass (from $0.65$ to $1.05$~\Msun, in steps of $0.05$~\Msun), the initial post-CE main-sequence mass (from $1.05$ to $1.30$~\Msun, in steps of $0.05$~\Msun), and the orbital period (from $1$ to $10$~d, in steps of $0.5$~d).
While both the fraction of mass lost from the WD and from the circumbinary disk were fixed, as described in Sect.~\ref{MESAassumptionsBinary}, we varied the fraction of mass lost from the vicinity of the donor star $\alpha$ (from $0$ to $0.15$, in steps of $0.05$).
We further varied the magnetic braking boosting factor ($\eta=1,10,100$), the mixing length ($1.5\,H_{\rm p}$, $1.0\,H_{\rm p}$, $2.5\,H_{\rm p}$), and the extent of the overshooting (from $0.015\,H_{\rm p}$ to $0.060\,H_{\rm p}$, in steps of $0.015\,H_{\rm p}$), where $H_{\rm p}$ is the pressure scale height at the convective boundary.
In each model, we adopted solar metallicity (i.e., ${Z=0.02}$) and assumed that the orbit was circular.

After finding the best-fitting post-CE model, we were able to search for the best-fitting pre-CE model, by using as constraints its initial parameters (i.e., the initial post-CE WD mass, donor star mass, and orbital period).
For simplicity, we assumed that the zero-age main-sequence binary orbits were circular.

We ran a large grid of binary models varying the zero-age mass of the hot WD progenitor from $2$ to $5$~\Msun, in steps of $0.01$~\Msun~and the zero-age orbital period from $10^3$ to $10^4$~d, in steps of $5$~d.
And, finally, the zero-age mass of the companion was chosen to be slightly smaller than the observed value as it increases during binary evolution due to wind accretion.
After identifying within the grid the model in which the properties of the binary just after CE evolution were closest to the initial best-fitting post-CE model, we refined the grid around this model to obtain the best-fitting pre-CE model.

\section{Formation pathway for \paradoxical} 
\label{Results}

With the fitting scheme described in Sect.~\ref{Search}, we found that a binary with zero-age main-sequence masses of ${\sim3.9}$ and ${\sim1.2}$~\Msun~and orbital period of ${\sim3800}$~d, evolves to the required initial post-CE binary to explain \paradoxical, which corresponds to a WD mass of ${\sim0.85}$~\Msun, a main-sequence mass of ${\sim1.25}$~\Msun, and an orbital period of ${\sim7}$~d.
In what follows we discuss in more detail how a zero-age main-sequence binary with these properties evolves to a binary having the properties of \paradoxical.

\begin{figure*}
\begin{center}
\includegraphics[width=0.99\linewidth]{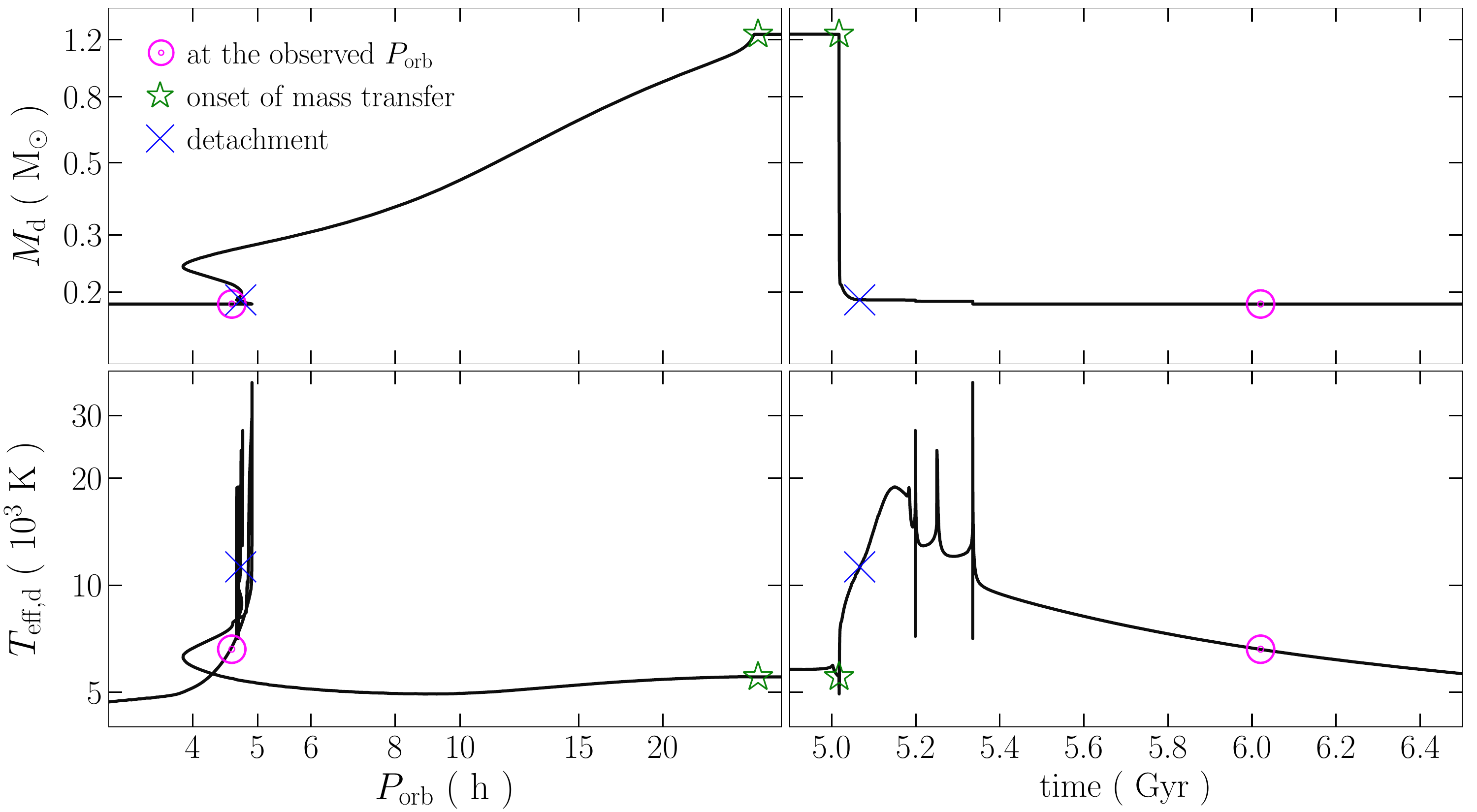}
\end{center}
\caption{Donor mass (top panels) and donor effective temperature (bottom panels) evolution with orbital period (left panels) and time (right panels). The magenta solar symbols correspond to the moment the orbital period is the same as observed, the green hollow stars mark the onset of mass transfer and the blue crosses indicate the end of the \cv~phase. More details are provided in Sect.~\ref{Results}.}
\label{Fig:M2Teff2}
\end{figure*}

\begin{figure*}
\begin{center}
\includegraphics[width=0.99\linewidth]{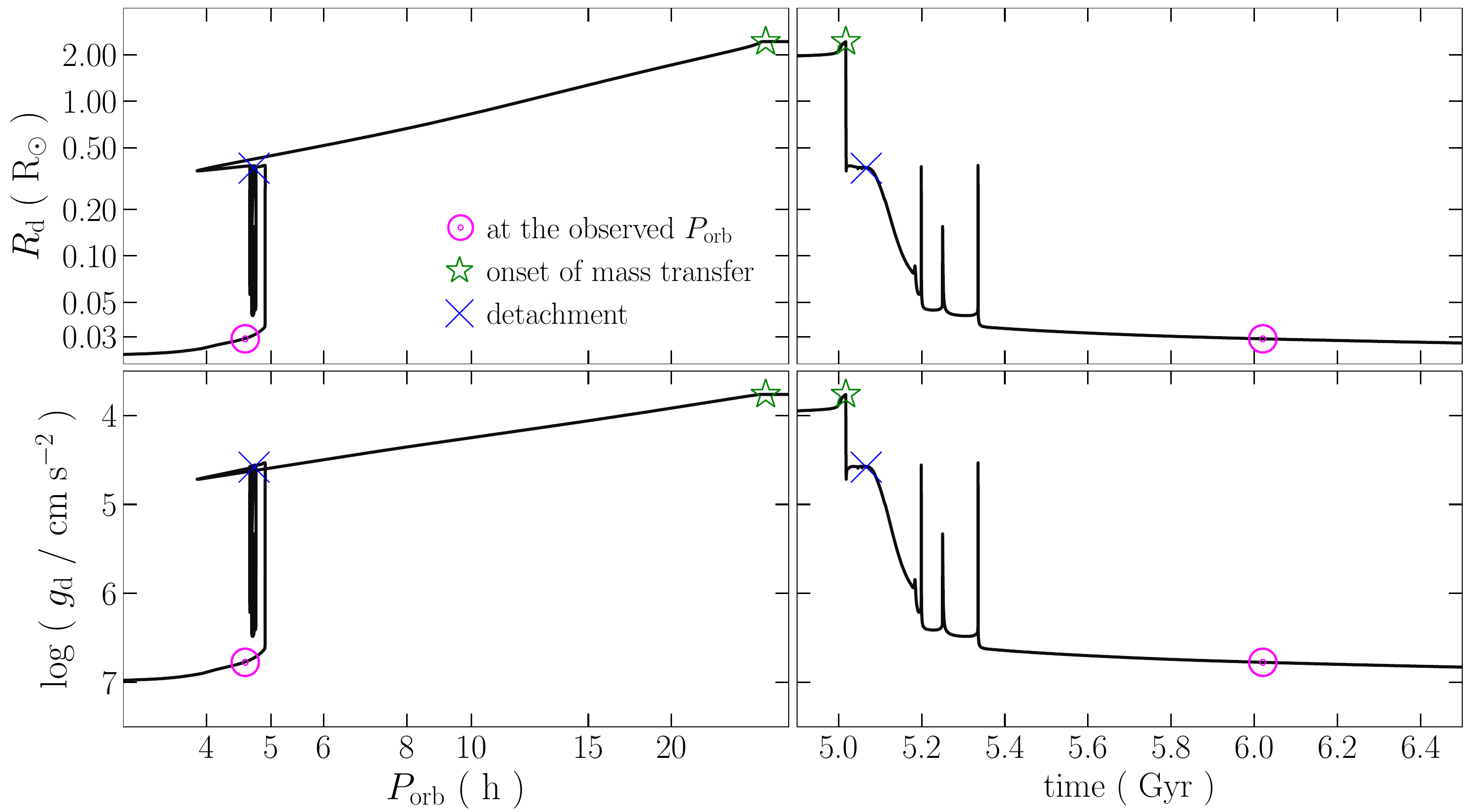}
\end{center}
\caption{Donor radius (top panels) and donor $\log{}g$ (bottom panels) evolution with orbital period (left panels) and time (right panels). The magenta solar symbols correspond to the moment the orbital period is the same as observed, the green hollow stars mark the onset of mass transfer and the blue crosses indicate the end of the \cv~phase. More details are provided in Sect.~\ref{Results}.}
\label{Fig:R2logg2}
\end{figure*}

\begin{figure*}
\begin{center}
\includegraphics[width=0.99\linewidth]{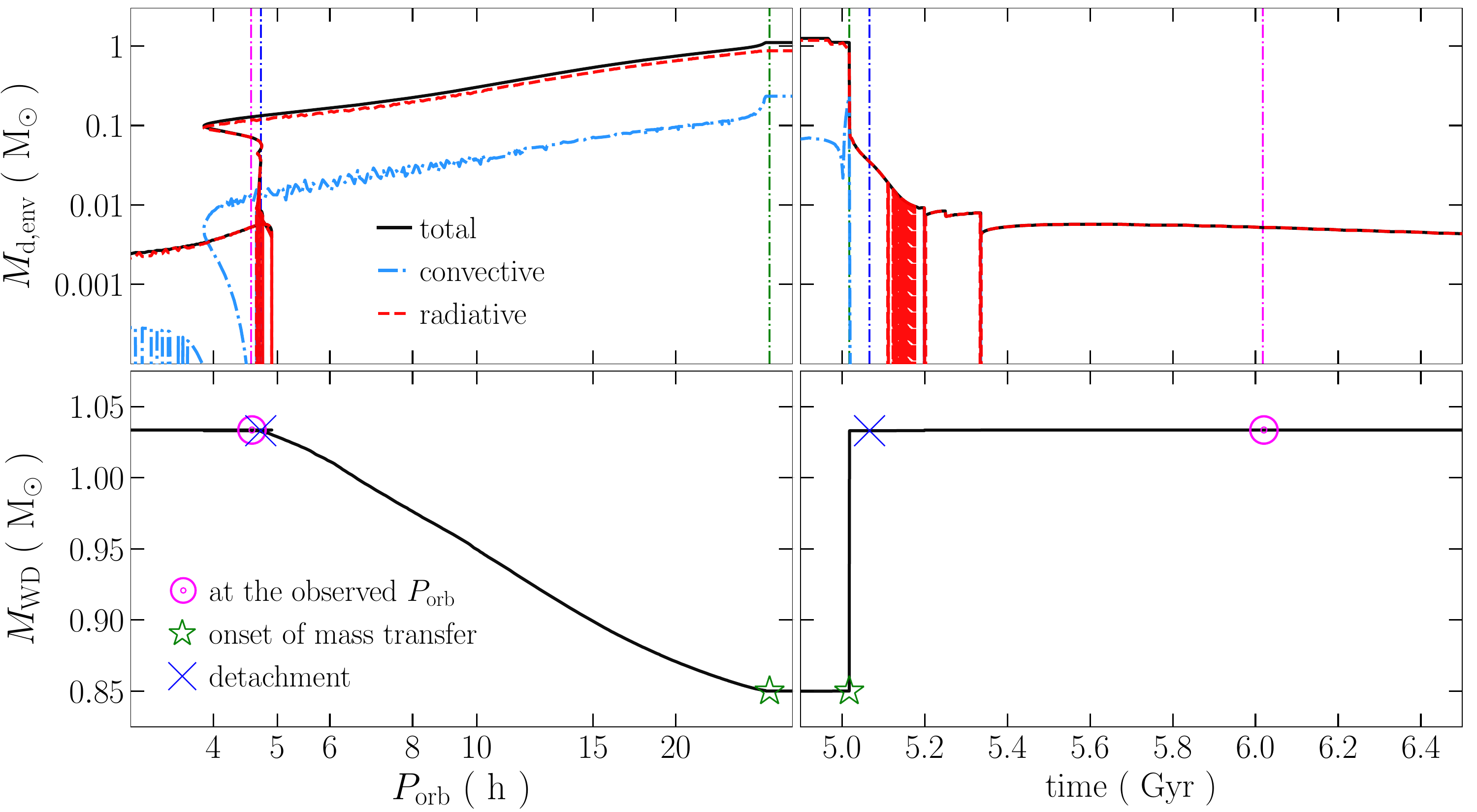}
\end{center}
\caption{Donor envelope mass (top panels) and WD mass (bottom panels) evolution with orbital period (left panels) and time (right panels). The magenta solar symbols and magenta vertical lines correspond to the moment the orbital period is the same as observed, the green hollow stars and green vertical lines mark the onset of mass transfer and the blue crosses and blue vertical lines indicate the end of the \cv~phase. More details are provided in Sect.~\ref{Results}.}
\label{Fig:Menv2Mwd}
\end{figure*}

\begin{figure*}
\begin{center}
\includegraphics[width=0.99\linewidth]{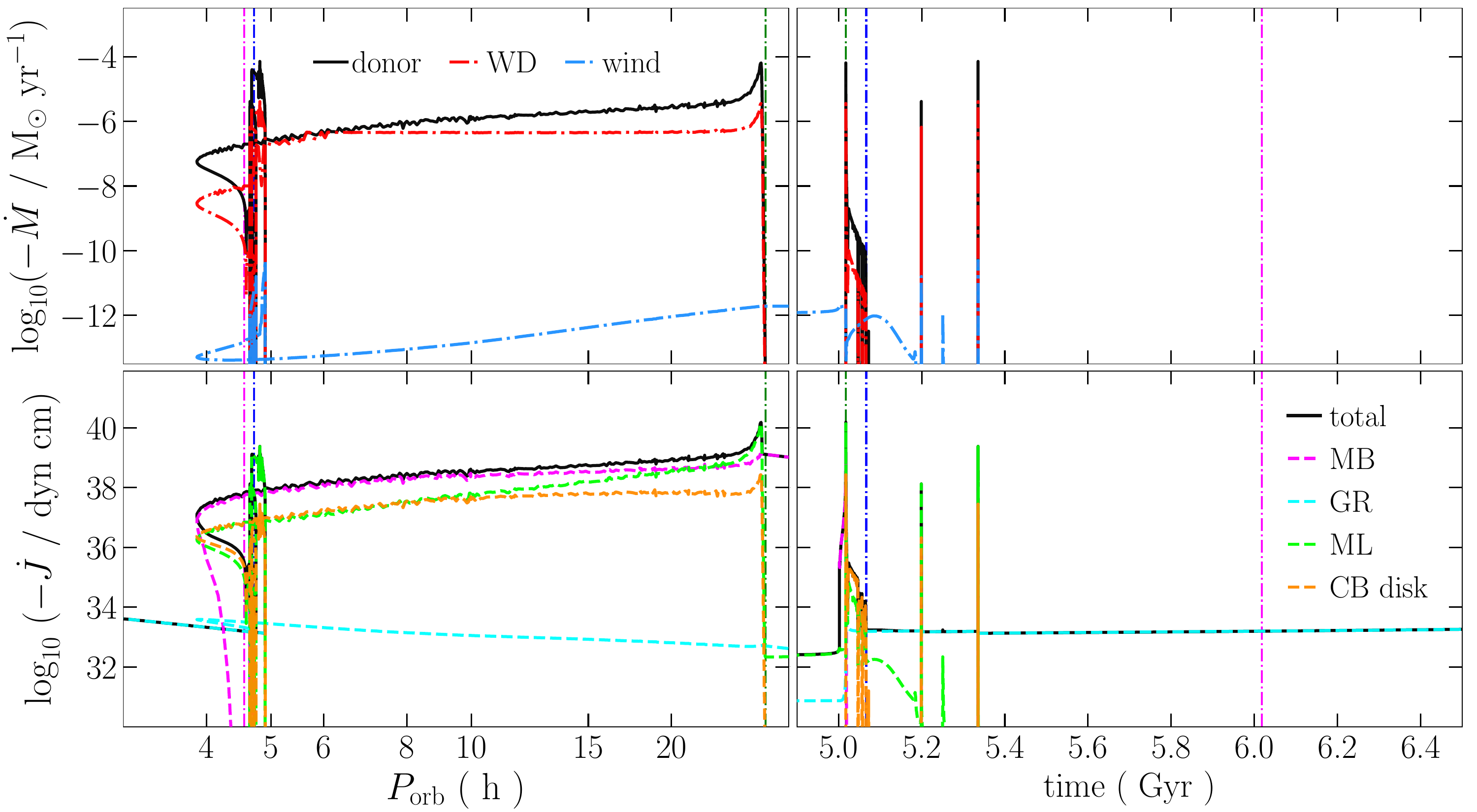}
\end{center}
\caption{Mass transfer rate, accretion rate, donor wind mass loss rate,  (top panels) and orbital angular momentum loss rate (bottom panels) evolution with orbital period (left panels) and time (right panels). The vertical blue lines indicate the time and orbital period at which the binary detaches, the vertical magenta lines correspond to the observed orbital period and the vertical green lines mark the onset of mass transfer. More details are provided in Sect.~\ref{Results}.}
\label{Fig:MdotAML}
\end{figure*}

\begin{figure*}
\begin{center}
\includegraphics[width=0.49\linewidth]{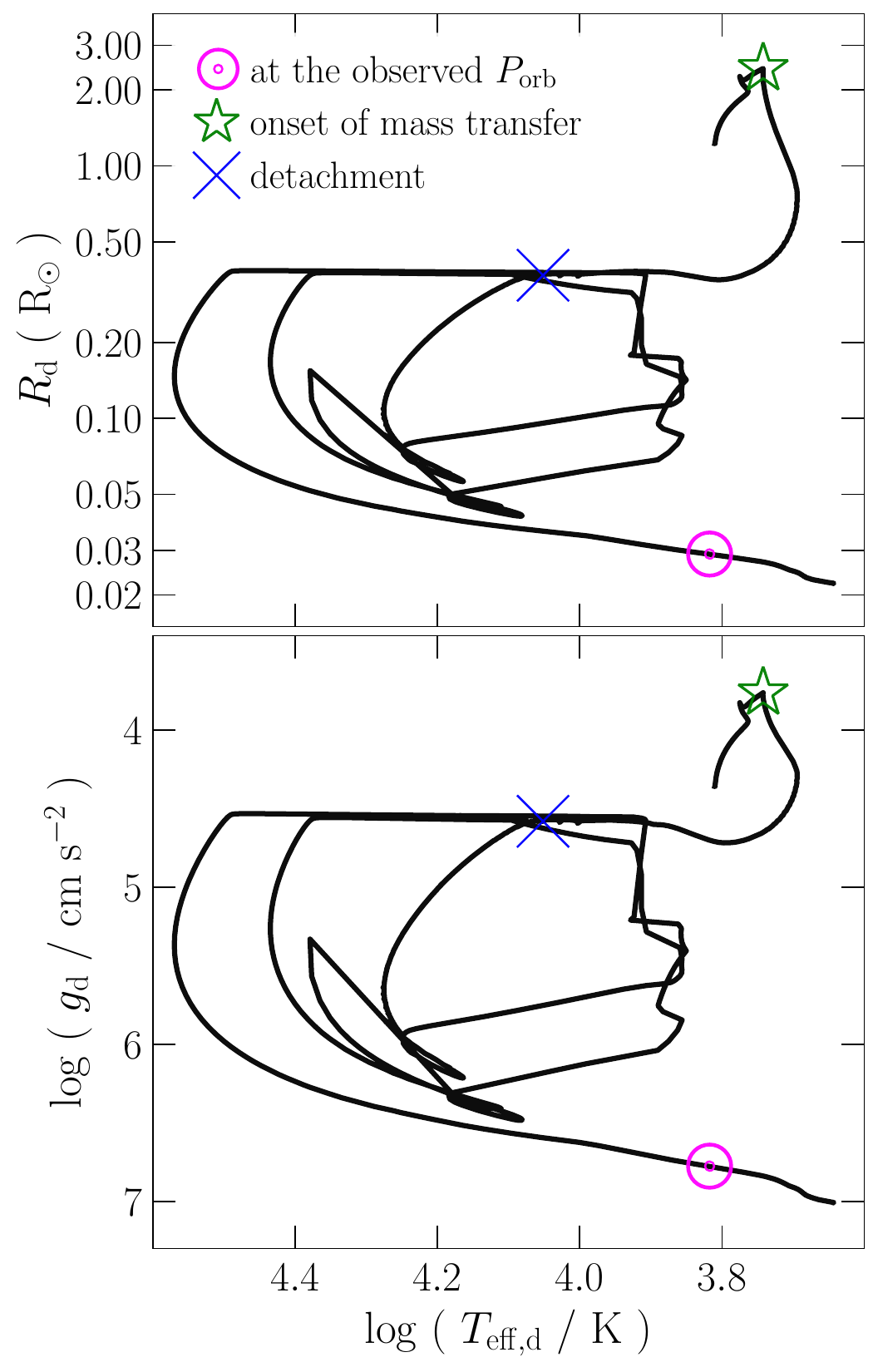}
\includegraphics[width=0.49\linewidth]{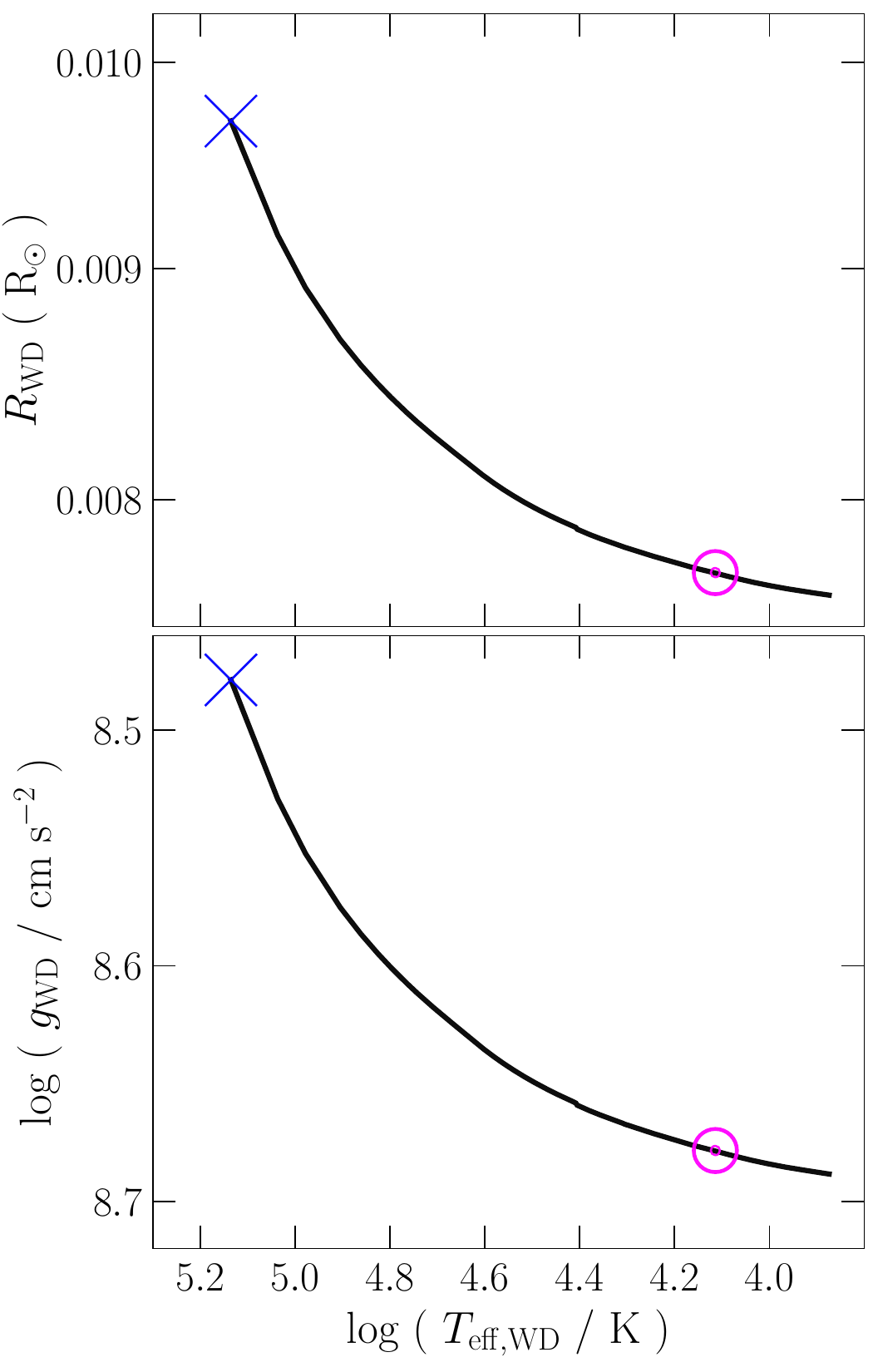}
\end{center}
\caption{Evolution of the massive WD (right panels) and the progenitor of the \elm~WD (left panels) in the planes radius versus effective temperature (top panels) and $\log{}g$ versus effective temperature (bottom panels). The magenta solar symbols correspond to the moment the orbital period is the same as observed, the green hollow stars mark the onset of mass transfer and the blue crosses indicate the onset of the detachment. More details are provided in Sect.~\ref{Results}.}
\label{Fig:HR}
\end{figure*}

The best-fitting parameters for the post-CE evolution with MESA are summarized in Table~\ref{Tab:Assumptions}, the sequence of events is given in Table~\ref{Tab:FormationChannel}, and the properties we predict in our modeling with MESA are compared with the observed ones in Table~\ref{Tab:Comparison}.
The post-CE evolution of the main properties of the binary and both components are shown in Figs.~\ref{Fig:M2Teff2}, \ref{Fig:R2logg2}, \ref{Fig:Menv2Mwd}, and \ref{Fig:MdotAML}, where we depict the evolution with orbital period and time of the donor mass, donor effective temperature, donor radius, donor $\log{}g$, donor envelope mass, WD mass, mass transfer rate, accretion rate, and orbital angular momentum loss.
We show in Fig.~\ref{Fig:HR} the radii and $\log{}g$ of both components as a function of their effective temperatures.

\begin{table}
\caption{Best-fitting MESA parameters for post-CE evolution.}
\label{Tab:Assumptions}
\centering
\setlength\tabcolsep{10.00pt} 
\renewcommand{\arraystretch}{1.4} 
\begin{tabular}{lc}
\hline
Parameter & Value \\
\hline
initial post-CE orbital period &  $7.07$~d \\
initial post-CE WD mass &  $0.85$~\Msun \\
initial post-CE companion mass &  $1.25$~\Msun \\
magnetic braking boosting factor ($\eta$) &  100\\
fraction of mass lost from WD ($\beta$)      &  $\leq85$\%\\
fraction of mass lost from donor ($\alpha$)  &  $15$\% \\
mixing length & $2.5\,H_{\rm p}$ \\
extent of exponential core overshooting & $0.06\,H_{\rm p}$ \\
\hline
\end{tabular}
\end{table}

\begin{table*}
\centering
\caption{Evolution of a zero-age main-sequence binary toward \paradoxical.}
\label{Tab:FormationChannel}
\setlength\tabcolsep{9pt} 
\renewcommand{\arraystretch}{1.25} 
\begin{tabular}{r c c c c r l}
\hline
\noalign{\smallskip}
 Time  &   $M_1$    &   $M_2$    & Type$_1$  & Type$_2$ & Orbital Period & Event\\
 (Myr) & (M$_\odot$)&(M$_\odot$) &           &          &  (days)        &      \\
\hline
\noalign{\smallskip}
   0.0000  &  3.920  &  1.204 & MS     & MS       & 3788.000  &  zero-age MS--MS binary \\
 188.5753  &  3.920  &  1.204 & SG     & MS       & 3788.104  &  change in primary type \\
 189.4994  &  3.920  &  1.205 & FGB    & MS       & 3788.482  &  change in primary type \\
 190.2362  &  3.919  &  1.205 & CHeB   & MS       & 3790.178  &  change in primary type \\
 225.0617  &  3.880  &  1.205 & E-AGB  & MS       & 3847.527  &  change in primary type \\
 226.6515  &  3.842  &  1.205 & TP-AGB & MS       & 3831.904  &  change in primary type \\
 227.1598  &  2.818  &  1.250 & TP-AGB & MS       & 3092.126  &  begin RLOF (primary is the donor) \\
 227.1598  &  2.818  &  1.250 & TP-AGB & MS       & 3092.126  &  CE evolution $(\alpha_{\rm CE}=0.3)$\\
 227.1598  &  0.850  &  1.250 & WD     & MS       &    7.070  &  end RLOF \\
5002.3338  &  0.850  &  1.246 & WD     & SG       &    7.092  &  change in secondary type \\
5017.5672  &  0.850  &  1.246 & WD     & SG       &    1.155  &  begin RLOF (secondary is the donor) \\
5066.7650  &  1.033  &  0.189 & WD     & proto-WD &    0.197  &  end RLOF \\
5335.6121  &  1.034  &  0.184 & WD     & WD       &    0.204  &  change in secondary type \\ 
\textbf{6017.6280}  &  \textbf{1.034}  &  \textbf{0.184} & \textbf{WD}     & \textbf{WD}       &    \textbf{0.190}  &  \textbf{binary looks like \paradoxical} \\
\noalign{\smallskip}
\hline
\end{tabular}
\begin{tablenotes}
\item[] \textbf{Notes.} For the pre-CE evolution and CE evolution we used the BSE code and for the post-CE evolution the MESA code. $M_1$ and $M_2$ and Type$_1$ and Type$_2$ are the masses and stellar types of the primary (progenitor of the hot WD) and secondary (progenitors of the cold WD), respectively. $P_{\rm orb}$ is the orbital period and the last column corresponds to the event occurring to the binary at the given time in the first column. The row in which the binary has the present-day properties of \paradoxical~is highlighted in boldface.
Abbreviations:
MS~(main-sequence~star),
SG~(subgiant~star),
FGB (first~giant~branch~star),
CHeB (core~helium~burning~star),
E-AGB (early~asymptotic~giant~branch~star),
TP-AGB~(thermally pulsing~asymptotic~giant~branch~star),
WD~(white~dwarf),
RLOF~(Roche~lobe~overflow),
CE~(common~envelope).
\end{tablenotes}
\end{table*}

\begin{table*}
\caption{Predicted and observed \citep{Bours_2015,ArosBunster_2025} parameters of \paradoxical.}
\label{Tab:Comparison}
\centering
\setlength\tabcolsep{8pt} 
\renewcommand{\arraystretch}{1.45} 
\begin{tabular}{lcccccc}
\hline
\vspace{-0.2cm}
\multirow{2}{*}{Parameter} & Observed & Observed              & Observed              & & \multirow{2}{*}{Predicted} \\
\vspace{0.1cm}
                           &          & (fixed ${\log{g}=6}$) & (fixed ${\log{g}=7}$) & & \\
\hline
$P_{\rm orb}$ (h) & $4.56$ & $4.56$ & $4.56$ & & $4.556$ \\   
$R_{\rm cold}/R_{\rm hot}$ & $4.27\pm0.09$ & $4.21\pm0.10$ & $3.89\pm0.09$ & & $3.767$ \\
\vspace{0.1cm}
total age (Gyr) & ${\gtrsim4}$ & ${\gtrsim4}$ & ${\gtrsim4}$ & & $6.017$ \\
\hline
 \multicolumn{6}{c}{\textit{hot WD}}  \\ 
$T_{\rm eff}$ (K)            & $13\,030\pm220$   & $12\,965\pm230$   & $12\,811\pm240$   & & $13\,117$ \\
$M$ (\Msun)                  & $1.06\pm0.05$     & $1.04\pm0.05$     & $1.00\pm0.05$     & & $1.034$ \\
$\log{}g$ (cm\,s$^{-2}$)     & $8.73\pm0.10$     & $8.70\pm0.12$     & $8.61\pm0.14$     & & $8.68$ \\
$R$ (\Rsun)                  & $0.0074\pm0.0006$ &                   &                   & & $0.0077$ \\
\vspace{0.1cm}
cooling age (Gyr)            & $1.0\pm0.1$       & $1.0\pm0.1$       & $0.9\pm0.1$       & & $0.951$ \\
\hline
\multicolumn{6}{c}{\textit{cold WD}}  \\  
$T_{\rm eff}$ (K)            & $6\,400\pm87$     & $6\,395\pm80$     &  $6\,460\pm80$    & & $ 6\,606$ \\
$M$ (\Msun)                  & ${\lesssim0.24}$  & ${\lesssim0.24}$  & ${\lesssim0.24}$  & & $0.184$ \\
$\log{}g$ (cm\,s$^{-2}$)     & ${5.26\pm0.36}$   & $6.0$             & $7.0$             & & $6.78$ \\
$R$ (\Rsun)                  & $0.032\pm0.003$   &                   &                   & & $0.029$ \\
\vspace{-0.2cm}
cooling age (Gyr) since      & \multirow{2}{*}{} & \multirow{2}{*}{} & \multirow{2}{*}{} & & \multirow{2}{*}{$0.951$} \\
\vspace{0.1cm}
the onset of the detachment  &  &  \\
\hline
\end{tabular}
\begin{tablenotes}
\item[] \textbf{Notes.} The properties of the cold WD were obtained from the MESA simulations, while those for the hot WD were obtained from the tracks computed by \citet[][thick atmosphere]{Bedard_2020}, assuming it was entirely rejuvenated during \cv~evolution (i.e., its age is zero at the detachment).
\end{tablenotes}

\end{table*}

Initially, the more massive star (${\approx3.92}$~\Msun) evolves off the main sequence and becomes a subgiant (at ${\approx188.58}$~Myr), and subsequently a red giant star (at ${\approx189.50}$~Myr), a central helium burning star (at ${\approx190.24}$~Myr), an early AGB star (at ${\approx225.06}$~Myr), and a thermally pulsing AGB star at ${\approx226.65}$~Myr.
Before filling its Roche lobe, the mass loss from the thermally pulsing AGB star reaches sufficiently high rates (up to ${\sim4\times10^{-5}}$~\Msunyr).
This makes wind accretion efficient enough so that the mass of the main-sequence star increases from ${\approx1.2}$ to ${\approx1.25}$~\Msun.
The thermally pulsing AGB star fills its Roche lobe triggering CE evolution quickly afterward (at ${\approx227.15}$~Myr) when it has a mass of ${\approx2.82}$~\Msun~and its hydrogen-free core has a mass of ${\approx0.85}$~\Msun.
After CE evolution, the binary hosts a WD and a main-sequence star with masses of ${\approx0.85}$~and ${\approx1.25}$~\Msun, respectively, and the orbital period is ${\approx7.07}$~d.

Just after the CE evolution, the companion of the WD is still a main-sequence star and has a convective core and radiative envelope, which means magnetic braking has not kicked in yet.
The binary then evolves toward longer periods due to the mass loss from the main-sequence star through stellar winds.
As soon as the central hydrogen abundance is substantially low (${\sim3\%}$) so that the central hydrogen burning rate is not enough to prevent the collapse of the star (at ${\approx4920.40}$~Myr), the overall contraction phase starts.
During this phase, the central hydrogen abundance drops to zero, the inert helium core becomes radiative, and a non-negligible portion of the envelope becomes convective.

When the temperature at the base of the hydrogen shell surrounding the inert helium core is sufficiently high, hydrogen shell burning starts and the overall contraction phase ends.
The star then becomes a subgiant (at ${\approx5002.33}$~Myr) and its core is entirely radiative.
Thus, the conditions for magnetic braking to act are satisfied (i.e., the core is radiative and a non-negligible portion of the envelope is convective) and from this point on it is allowed to extract orbital angular momentum.
The binary then evolves toward shorter periods.
Toward the onset of mass transfer, the rate at which orbital angular momentum is lost due to magnetic braking increases from ${\sim10^{35}}$ to ${\sim10^{39}}$~dyn~cm.

The subgiant fills its Roche lobe when the orbital period is ${\approx1.16}$~d (at ${\approx5017.57}$~Myr).
Just after the onset of mass transfer, the mass loss from the binary drives the evolution but quickly afterward magnetic braking becomes the main driver throughout \cv~evolution.
During \cv~evolution, the orbital period and the donor mass quickly decrease as a result of the high mass transfer rates (${\sim10^{-6}-10^{-5}}$~\Msunyr) due to the strong magnetic braking torques (${\sim10^{38}-10^{39}}$~dyn~cm).
As a consequence, the radius of the donor decreases, while its $\log{}g$ increases.
Meanwhile, the donor effective temperature is roughly constant (${\sim5000-5500}$~K).
As the mass accretion rate is sufficiently high so that stable hydrogen burning occurs, the WD mass increases to ${\approx1.03}$~\Msun~and its age is reset (i.e., the WD rejuvenates).

Toward the end of the \cv~evolution (at ${\approx5018.46}$~Myr), when the donor mass is ${\approx0.24}$~\Msun~and the orbital period is ${\approx3.87}$~hr, the mass fraction of the convective envelope becomes sufficiently small (${\approx2}$\%).
From this point on, the envelope of the donor is virtually entirely radiative and magnetic braking gets significantly reduced.
As the orbital angular momentum loss rates due to mass loss ({$\sim10^{36}$}~dyn~cm) and gravitational radiation ({$\sim10^{34}$}~dyn~cm) are much weaker than magnetic braking before disruption ({$\sim10^{37}$}~dyn~cm), they are not sufficiently strong to make the evolution convergent.
The orbital period then starts to increase as a response to the nuclear evolution of the donor.
The effective temperature and the radius of the donor increase, while its $\log{}g$ decreases in response.

At ${\approx5066.77}$~Myr, the binary detaches with an orbital period of ${\approx4.72}$~hr.
The donor becomes an \elm~proto-WD having a mass of ${\approx0.19}$~\Msun, an envelope mass of ${\approx0.035}$~\Msun, and an effective temperature of ${\approx11\,258}$~K.
From this point on, the newly formed proto-WD undergoes a few hydrogen flashes for ${\sim300}$~Myr.
After this phase (at ${\approx5335.63}$~Myr), the proto-WD finally enters its cooling sequence (i.e., becomes an \elm~WD).
At the present day (i.e., at ${\approx6017.63}$~Myr), the orbital period and the properties of both \elm~WD and massive WD are consistent with the observed.

\section{Discussion}

\subsection{Impact of assumptions and model parameters} 
\label{ModelParameters}

We found in the previous sections a reasonable model for \paradoxical.
In our fitting scheme, several parameters were assumed while others were fitted.
We discuss below the influence of these assumptions and fitted parameters.

\subsubsection{White dwarf rejuvenation}

The most important assumption we made is that the massive WD highly/completely rejuvenates during \cv~evolution, which means its age is reset to zero at the detachment.
The properties of an accreting WD are likely to change due to the heating provided by the compression of the incoming matter.
In particular, the release of gravitational energy (i.e., conversion into thermal energy) during the compression of the accreted material is expected to heat its interior \citep[e.g.,][]{Sion_1995,Epelstain_2007,Townsley_2004,Townsley_2009}.
This energy is subsequently transported outward to the surface and inward to the core.
This would then cause an overall rejuvenation of the WD.

According to the evolutionary models calculated by \citet{Bedard_2020}, the core and effective temperature of the 0.85-\Msun~WD at the onset of \cv~evolution are ${\sim3\times10^6}$ and ${\sim6\times10^3}$~K, respectively.
At this moment, the 0.85-\Msun~WD is ${\sim4.79}$~Gyr old and is highly crystallized (${\sim78}$\%).
During \cv~evolution, the effective temperature of the WD increases to  ${\gtrsim10^5}$~K, according to \citet[][their Eq.~2]{Townsley_2009}, due to the high mass transfer rates (${\sim10^{-7}-10^{-6}}$~\Msunyr) onto the WD.
This heating would then travel inward and should be enough to heat the core, causing an overall rejuvenation of the WD.

We should stress that this effective temperature we estimated is most likely wrong because \citet[][]{Townsley_2009} derived a relation between effective temperature due to compressional heating and the mass transfer rate for accreting WDs undergoing unstable hydrogen burning, which is not the case in our model during most of the evolution.
To the best of our knowledge, unfortunately, there is no dedicated modeling (or even theory) available for WDs accreting at such high mass transfer rates and undergoing stable hydrogen burning.
That being said, although our assumption for rejuvenation seems reasonable, it is still unclear how the deepness and the timescale for the heat traveling inward depend on the model parameters. 
Therefore, it remains to be verified under which conditions this idea could work.

\subsubsection{Orbital angular momentum loss}

The most important fitted parameter in our fitting scheme is the boosting factor of the CARB model, required to explain the properties of both WDs and the orbital period.
To reproduce the age, $\log{}g$, and effective temperature of the rejuvenated massive WD, the \elm~proto-WD evolution needs to be shorter than a few hundred megayears.
The evolution of an \elm~proto-WD can be speeded up if it develops weak flashes as a result of thermal instabilities.
These flashes cause the hydrogen envelopes to be thinner, which prevents stable nuclear burning from being a sizeable energy source at advanced
stages of evolution \citep[e.g.,][]{Althaus_2001,Panei_2007}.
Because of that, the cooling age of the \elm~proto-WD becomes much shorter, with lifetimes of only a few hundred megayears.
Thus, a required condition to explain \paradoxical~in the scenario we propose is that the \elm~proto-WD undergoes hydrogen shell flashes.

In our best-fitting model, the CARB magnetic braking is enhanced by a factor of ${\sim100}$.
This boosting is required to make the \cv~evolution strongly convergent during the formation of the an \elm~proto-WD that undergoes hydrogen shell flashes, as otherwise the observed orbital period cannot be reproduced for this type of proto-WD.
In the case the standard CARB model is assumed, or the boosting factor is smaller than ${\sim100}$, the resulting orbital period of the double WD binary is always longer than observed.
This remains true regardless of the assumed strength of other sources of orbital angular momentum loss.

In our fitting scheme, we allowed other sources of orbital angular momentum loss, namely mass loss from the vicinities of the WD and the donor and mass loss through the outer Lagrange point (circumbinary disk).
We found that the mass of the massive WD could only be reproduced if a non-negligible fraction (${\sim15}$\%) of the mass that would be transferred from the donor to the WD leaves the binary and if the WD retains at least a non-negligible portion of the accreted mass (${\sim15}$\%) regardless of the mass transfer rate.
However, it is very likely that the mass of the massive WD can be fitted by different combinations of these mass loss efficiencies.

Most importantly, these orbital angular momentum losses originating from the mass transfer process itself, that is, the so-called consequential angular momentum loss, cannot contribute to reducing the strength of magnetic braking since the mechanisms responsible for consequential angular momentum losses cannot be the main driver of the evolution.
This happens because there is a maximum rate at which angular momentum is extracted from the orbit during dynamically stable mass transfer.
In the case more orbital angular momentum is lost than this maximum rate, a runway process will be triggered making mass transfer dynamically unstable and subsequently leading to a merger.
For instance, in all our attempts to make the orbital angular momentum loss due to a circumbinary disk as strong as required to explain the observed orbital period, mass transfer became dynamically unstable.
This happened because consequential angular momentum loss leads to positive feedback, that is, the larger the disk mass, the stronger the torque (i.e., the higher the orbital angular momentum loss rate) and in turn the higher the mass transfer rate.
At a certain critical mass transfer rate, a runaway process is triggered for higher mass transfer rates.
In the particular case of \paradoxical, the required mass transfer rates to explain their properties are higher than such a critical rate.
Therefore, consequential orbital angular momentum loss has in general a minor impact on the evolution, except at the very beginning and the very end.

We would like to stress that we incorporated element diffusion due to gravitational settling and chemical and thermal diffusion into the calculations.
As shown by \citet{Istrate_2016}, there is a minimum proto-WD mass above which hydrogen shell flashes occur.
Such a minimum mass is ${\sim0.21}$~\Msun~for models without diffusion enabled but it drops to ${\sim0.17}$~\Msun~when diffusion is included.
This happens because there is a tail in the hydrogen distribution that chemically diffuses inward where the temperature is high enough to burn it, triggering a thermal runaway for lower masses \citep{Althaus_2001}.
As the helium core mass correlates with the degree of nuclear evolution of a subgiant star, \cv~evolution without element diffusion would need to take place with a more nuclear evolved donor.
In this case, the magnetic braking would need to be even stronger than we found while adopting element diffusion to make the evolution convergent.

\subsubsection{Mixing and overshooting}

Other less important parameters for the success of our modeling include the mixing length and the extent of the core overshooting.
For a fixed initial donor mass, the change of its radius and the growth of its helium core mass during main-sequence and subgiant phases are affected by the mixing length and the extent of the core overshooting, respectively.
In particular, the larger the mixing length, the smaller the radius, and the larger the extent of the core overshooting, the higher the helium core mass.
We found that the characteristic of \paradoxical~can be explained if the mixing length is ${\sim2.5\,H_{\rm p}}$ and the extent of the core overshooting is ${\sim0.06\,H_{\rm p}}$, which cause the onset of mass transfer to take place at a sufficiently short orbital period and the mass of the helium core of the donor to be sufficiently high.

The mixing length we obtained is larger than typically adopted values of this parameter \citep[$\sim2$; e.g.,][]{Claret_2019,Viani_2020,Joyce_2023}.
However, it is very unlikely that the mixing length should be unique for all stars, irrespective of their masses, evolutionary phases, and metallicity \citep[e.g.,][]{Creevey_2017,Joyce_2023}.
In addition, as shown by \citep[e.g.,][]{Valle_2019}, the standard mixing-length calibrations can result in values with a large spread, being on average ${\sim2.20\pm0.52}$.
That being said, we believe our best-fitting mixing length is reasonable.
The extent of the core overshooting is a bit more complicated to compare to observationally derived values because the employed method may not be reliable  \citep[e.g., eclipsing binaries][]{ConstantinoBaraffe_2018} and there are several treatments available for core overshooting \citep[e.g.,][]{Anders_2023}.
Bearing that in mind, our extent is higher (by a factor of at least ${\sim2}$) than asteroseismically derived extents of pulsating main-sequence stars \citep[][their fig.~12e]{Anders_2023}.
Despite that, considering the underlying uncertainties in constraining the amount of overshoot, we believe our higher-than-average extent does not invalidate the scenario we propose.

\subsubsection{Total age}

To reproduce the total age of the triple (i.e., ${\gtrsim4}$~Gyr), the zero-age mass of the progenitor of the \elm~WD has to be ${\lesssim1.3}$~\Msun~(for a metallicity of ${Z=0.02}$), which implies a pre-CV evolution time of ${\gtrsim3}$~Gyr.
\Cv~evolution is very quick due to the strong orbital angular momentum loss making the onset of the detachment (i.e., the moment at which the \elm~proto-WD is formed) at ${\gtrsim3}$~Gyr.
The massive WD and its companion then evolve for ${\sim1}$~Gyr to the present day.

Thus, for higher zero-age masses of the progenitor of the \elm~WD, the present-day total age of \paradoxical~would be smaller than ${\sim4}$~Gyr.
We found that an initial post-CE mass of ${\sim1.25}$~\Msun~for the progenitor of the \elm~WD leads to a total age of ${\sim6}$~Gyr, consistent with observations.
To achieve older (younger) total ages, we would simply need to assume a lower (higher) initial post-CE mass.

\subsubsection{Additional considerations}

We would like to draw the readers attention to the fact that the combination of model parameters we found in our fitting scheme is most likely not the only one that can explain \paradoxical.
It is actually very likely that the properties of \paradoxical~can be reproduced by different combinations of mass loss efficiency, mass accretion efficiency, wind mass loss rates, wind mass accretion rates, extent of core overshooting, mixing length, zero-age properties and initial post-CE properties.

However, as we mentioned earlier, the zero-age (or initial post-CE) mass of the progenitor of the \elm~WD is relatively well constrained by the total age of the triple.
The orbital angular momentum loss rates are also relatively well constrained by the present-day properties of both WDs and the present-day orbital period.
In addition, to reproduce the present-day mass of the massive WD, the initial post-CE WD mass needs to be significantly smaller than that since during \cv~evolution its mass increases due to the required high mass transfer rates. 
Furthermore, the zero-age mass of the progenitor of the \elm~WD needs to be smaller than the required initial post-CE mass as its mass is likely increased due to wind mass accretion prior to the CE evolution.
That being said, we should also stress that despite the fact the model parameters can change, the overall formation pathway we propose here is thus far the most likely, as we discuss in Sect.~\ref{OtherFormationChannels}.

\subsection{Strength of magnetic braking in different populations} 
\label{StrengthMB}

We adopted the CARB model in this paper and found that magnetic braking needs to be very strong (${\sim10^{38}-10^{39}}$~dyn~cm) to explain \paradoxical.
These rates are much higher than those expected to explain single stars and other types of binaries.
We discuss below how this compares to the strength and the form of magnetic braking typically used to explain different populations.

Regarding detached binaries with low-mass main-sequence stars, \citet{ElBadry_2022} showed that the characteristics of detached binary main-sequence stars are better explained by a magnetic braking recipe that takes into account the saturation of the magnetic field.
These authors found that torques resulting in orbital angular momentum loss rates on the order of ${\sim10^{32}-10^{34}}$~dyn~cm are enough to reproduce their orbital period distributions.
Afterward, \citet{Belloni_2024a} showed that a saturated, boosted and disrupted magnetic braking could explain not only binary main-sequence stars but also the distribution of the fraction of detached post-CE binaries with WDs paired with M dwarfs across the M dwarf mass.
For that, the boosting factor should be such that the angular momentum loss rates are at least ${\sim10^{35}-10^{36}}$~dyn~cm for M dwarfs with radiative core and the disruption factor such that the rates are ${\sim10^{32}-10^{33}}$~dyn~cm for fully convective M dwarfs.
More recently, \citet{Blomberg_2024} showed that detached post-CE binaries with hot subdwarfs orbiting M dwarf are also better reproduced by a saturated, boosted and disrupted magnetic braking.

The saturated, boosted and disrupted magnetic braking prescription was also applied to standard \cvs~(i.e., those with unevolved donors) by \citet{BarrazaJorquera_2025}.
These authors showed that similar orbital angular momentum loss rates (i.e., ${\sim10^{35}-10^{36}}$~dyn~cm) can explain the orbital period gap \citep{Schreiber_2024}.
These rates are consistent with those in previous calculations with the so-called RVJ \citep*{RVJ} prescription, such as those carried out by \citet[][]{Knigge_2011_OK}.

Populations of compact objects accreting from nuclear evolved donors are in general better explained by stronger magnetic braking.
For instance, \citet{CARB}, who invented the CARB model, showed that stronger magnetic braking is required to explain persistent low-mass X-ray binaries.
The same prescription was adopted by \citet{BelloniSchreiber_2023}, who showed that the characteristic of \cvs~can be reasonably well explained.
A stronger magnetic braking is also required to solve the fine-tuning problem related to the formation of close detached binaries hosting millisecond pulsars with \elm~WDs companions \citep{Istrate_2014,Chen_2021,Soethe_2021}, which are descendants of low-mass X-ray binaries.
Not surprisingly, a stronger magnetic braking can also solve the fine-tuning problem associated with the formation of AM\,CVn binaries from \cvs~\citep{BelloniSchreiber_2023}.

We here found that the torques predicted by the CARB model are not strong enough to explain \paradoxical.
For a successful fit, we needed to assume a boosted version of the CARB recipe.
A useful extension of this work would be to fit the systems discovered by \citet{ElBadry_2021} in a similar detailed fashion we did for \paradoxical~as this could shed more light on the required strength of magnetic braking to fit the population of \cvs~with nuclear evolved donor.
Another interesting follow-up exercise would be to verify under which conditions the saturated, boosted and disrupted magnetic braking recipe would lead to similarly satisfactory evolution to explain the characteristics of \paradoxical~and \cvs~with nuclear evolved donors.
It would also be interesting to apply the saturated, boosted and disrupted magnetic braking to low-mass X-ray binaries and progenitors of AM\,CVn binaries.

\subsection{On the existence of alternative formation pathways} 
\label{OtherFormationChannels}

We argue in Sect.~\ref{Introduction} that the main formation channels tested for \paradoxical~does not work.
A formation pathway involving two episodes of CE evolution, having the massive WD formed first followed by the formation of the \elm~WD, does not work because the more massive WD would be much colder than its \elm~WD companion.
In addition, it is very unlikely that the binary would survive the second episode of CE evolution since the progenitor of the \elm~WD would be at best a red giant star descending from a solar-type star, which has a strongly bound envelope that cannot be easily ejected.
A formation channel in which the \elm~WD is formed first through dynamically stable mass transfer, followed by the formation of the massive WD via CE evolution, also does not work because the massive WD would be formed quickly after the \elm~WD leading to an \elm~WD that is much hotter than observed \citep{ArosBunster_2025}.

As an alternative to these standard formation channels, \citet{Jiang_2018} proposed a "strange" scenario for \paradoxical~in which the estimated different cooling ages of both WDs originate from the heating process during the formation of a strange dwarf.
These authors assumed that if the central density of a WD exceeds a critical density, a strange quark core would emerge in the central parts and the WD would evolve into a strange dwarf. 
In this case, during this transformation, the mass of the massive WD in \paradoxical~would decrease and the transport of energy would heat the WD, resulting in a hotter WD.
However, in the calculations by \citet{Jiang_2018}, the effective temperature of the \elm~WD can only be reproduced for a metallicity of ${Z\lesssim10^{-4}}$, which is inconsistent with the distance of \paradoxical~\citep[i.e., ${\sim120}$~pc,][]{BailerJones_2021} and its location (Milky Way disk).
We therefore believe that this scenario seems to be very unlikely.

\section{Conclusions}
\label{conclusions}

We ran binary evolution models using the \mesa~code in a bid to model \paradoxical~assuming the \cv~channel.
We adopted a boosted version of the CARB model for magnetic braking, which provides sufficiently high orbital angular momentum loss rates that \cv~evolution with more nuclear evolved donors is convergent, leading to double WDs in tighter orbits.
We find that \paradoxical~can be understood as being a descendant of a \cv~with an evolved donor, resolving the paradox.
In this scenario, the progenitor of the \elm~WD would have initially been a solar-type star that evolved into a subgiant before mass transfer began.
During \cv~evolution, the more massive WD would have undergone significant rejuvenation and the magnetic braking torques would need to be around $100$ times stronger than those predicted by the CARB model.
Our results further support the idea that \cvs~most likely play an important role in the formation of AM\,CVn binaries and close double WDs in general.

\begin{acknowledgements}

We would like to thank an anonymous referee for constructive comments and suggestions that helped to improve this manuscript.
DB acknowledges partial support from FONDECYT (grant number 3220167) and partial support from the São Paulo Research Foundation (FAPESP), Brazil, Process Numbers {\#2024/03736-2} and {\#2025/00817-4}.
MRS acknowledge financial support from {FONDECYT} (grant number {1221059}).

\end{acknowledgements}

\bibliographystyle{aa}
\bibliography{references}

\end{document}